\documentclass[aps,prl,twocolumn,showpacs]{revtex4}
\usepackage{txfonts}
\usepackage{graphicx}

\begin{document}

\title{Nodal Gap in Fe-Based Layered Superconductor LaO$_{0.9}$F$_{0.1-\delta}$FeAs Probed by Specific Heat Measurements}

\author{Gang Mu, Xiyu Zhu, Lei Fang, Lei Shan, Cong Ren and Hai-Hu Wen}\email{hhwen@aphy.iphy.ac.cn }

\affiliation{National Laboratory for Superconductivity, Institute of
Physics and Beijing National Laboratory for Condensed Matter
Physics, Chinese Academy of Sciences, P.O. Box 603, Beijing 100190,
People's Republic of China}

\begin{abstract}
We report the specific heat measurements on the newly discovered
Fe-based layered superconductor LaO$_{0.9}$F$_{0.1-\delta}$FeAs with
the onset transition temperature $T_c \approx 28$ K. A nonlinear
magnetic field dependence of the electronic specific heat
coefficient $\gamma(H)$ has been found in the low temperature limit,
which is consistent with the prediction for a nodal superconductor.
The maximum gap value $\Delta_0\approx$ 3.4$\pm$0.5 meV was derived
by analyzing $\gamma(H)$ based on the d-wave model. We also detected
the electronic specific heat difference between 9 T and 0 T in wide
temperature region, a specific heat anomaly can be clearly observed
near $T_c$. The Debye temperature of our sample was determined to be
about 315.7 K. Our results suggest an unconventional mechanism for
this new superconductor.

\end{abstract} 
\maketitle

In the superconducting state, the conduction electrons pair and
condense into a low energy state leading to the formation of a gap
at the Fermi energy $E_F$. It is this gap that protects the
superconducting condensate. Based on the general space group of the
material, and together with the detailed pairing interaction between
the two electrons, the superconducting gap should have a specific
symmetry for an individual superconductor. This gap symmetry is very
important for understanding the underlying mechanism of a
superconductor. For example, a $d_{x^2-y^2}$ symmetry has been
proved by tremendous experiments for the cuprate
superconductors.\cite{Tsuei1}. One effective way to detect the gap
symmetry is to generate quasiparticles (QP) from the condensate and
then trace out the detailed way for the accumulation of the QP
density of states (DOS). Specific heat is one of the powerful tools
to measure the DOS at the Fermi level. Textbook knowledge tells us
that, the low temperature electronic SH for an s-wave gap should
have an exponential temperature dependence, namely $C_e\propto
exp(-T_0/T)$, here $T_0$ is a characteristic temperature related to
the magnitude of the energy gap. For a superconductor with nodal gap
symmetry, the DOS is a power law of energy leading to a power law
dependence of temperature for the electronic SH: $C_e\propto T^2$
for the gap with line nodes and $C_e\propto T^3$ for point
nodes.\cite{Sigrist} For example, in cuprate superconductors, there
are line nodes in the gap function, this results in\cite{Kopnin1996}
an electronic SH $C_e=\alpha T^2$, where $\alpha \propto
\gamma_n/T_c$ and $\gamma_n$ is the specific heat coefficient
reflecting the DOS at the Fermi level of the normal state. In the
mixed state, the magnetic vortices will induce depairing both within
and outside the vortex cores leading to the localized and
delocalized QP DOS, respectively. Volovik \cite{Volovik} pointed out
that for d-wave superconductors in the mixed state, supercurrents
around a vortex core lead to a Doppler shift to the quasi-particle
excitation spectrum, which affects strongly the low energy
excitation around the nodes. It was shown that the contribution from
the delocalized part will prevail over the core part and the SH is
predicted to behave as\cite{Volovik,Kopnin1996}
$C_{vol}=k\gamma_nT\sqrt{H/H_{c2}}$ with $k$ in the order of unity,
$H_{c2}$ the upper critical field. This prediction has been verified
by many measurements which were taken as the evidence for d-wave
symmetry in cuprate
superconductors\cite{Moler,Revaz,Wright,Phillips,Nohara,Chen,WenHH,Hussey}.

In past years, superconductivity at several Kelvins was observed in
some quaternary oxypnictides with a general formula as LnOMPn (where
Ln= La and Pr; M = Mn, Fe, Co and Ni; Pn = P and
As)\cite{LOMP1,LOMP2,LOMP3}. It was found that the $T_c$ can be
increased by partially substituting the element $O$ with
$F$\cite{LOMP2}. Recently, it was reported that the superconducting
transition temperature can be increased to 26 K in the material
La[O$_{1-x}$F$_x$]FeAs ($x=0.05\sim0.12$)\cite{FirstJACS}. This is
surprising since the iron elements normally give rise to magnetic
moments, and in many cases they form a long range ferromagnetic
order, and are thus detrimental to the superconductivity with
singlet pairing. Therefore it is highly desired to know the gap
symmetry of this new superconductor. In this Letter, we report the
measurements on low temperature specific heat under different
magnetic fields. A nonlinear field dependence of SH coefficient
$\gamma$ is discovered. Our data together with a detailed analysis
indicate that the new superconductor $LaOFeAs$ may have a nodal gap.

The polycrystalline samples were synthesized by using a two-step
solid state reaction method. First the starting materials Fe powder
(purity 99.95\%) and As grains (purity 99.99\%) were mixed in 1:1
ratio, ground and pressed into a pellet shape. Then it was sealed in
an evacuated quartz tube and followed by burning at 700$^\circ$C for
10 hours. Then the resultant pellet was smashed and ground together
with the LaF$_3$ powder (purity 99.95\%), La$_2$O$_3$ powder (purity
99.9\%) and grains of La (purity 99.99\%) in stoichiometry as the
formula LaO$_{0.9}$F$_{0.1-\delta}$FeAs. Again it was pressed into a
pellet and sealed in an evacuated quartz tube and burned at about
940$^\circ$C for 2 hours, followed by a burning at 1150$^\circ$C for
48 hours. Then it was cooled down slowly to room temperature. Since
a little amount of F may escape during the second step fabrication,
in the formula for our sample, we use ($0.1-\delta$) as the possible
concentration of F. X-ray diffraction patterns in this sample show
that the dominant component is from LaO$_{0.9}$F$_{0.1-\delta}$FeAs.

The ac susceptibility were measured based on an Oxford cryogenic
system (Maglab-Exa-12). The resistivity and the specific heat was
measured on the Quantum Design instrument physical property
measurement system (PPMS) with temperature down to 1.8 K. We
employed the thermal relaxation technique to perform the specific
heat measurements. For the SH measurements, we used a latest
improved SH measuring puck from Quantum Design, which has negligible
field dependence of the sensor of the thermometer on the chip as
well as the thermal conductance of the thermal linking wires.

In the main frame of Fig. 1 we show the raw data of specific heat
plotted as $C/T$ vs $T^2$ under two different magnetic fields 0 T
and 9 T. Very similar to the case in cuprate superconductors, no
visible superconducting jump can be seen from the raw data at zero
field, indicating a rather small superfluid density or condensation
energy in the present system. By extrapolating the data in the
superconducting state at zero field down to 0 K, it is surprisingly
to see a rather small residual term $\gamma_0\approx$ 0.69 $mJ/mol
K^2$, indicating a very small non-superconducting volume or little
impurity scattering centers in our present sample. [Here one mole
means one unit cell or two molecules,
i.e.,(LaO$_{0.9}$F$_{0.1-\delta}$FeAs)$_2$ ] Inset (a) of Fig. 1
shows the temperature dependence of resistivity at 0 T in a wide
temperature range up to 300 K. The onset transition temperature is
about 28 K. One can also see that the residual resistivity ratio
(RRR) is about 18.5, which is a rather large value for a
polycrystalline sample. The residual resistivity at 30 K is only
0.13 m$\Omega$ cm, which is several times smaller than the reports
by other groups\cite{FirstJACS}. All these suggest that our samples
are much cleaner with fewer scattering centers or smaller
non-superconducting volume. The ac susceptibility data measured at
zero dc magnetic field were shown in the inset (b) of Fig. 1.

\begin{figure}
\includegraphics[width=8cm]{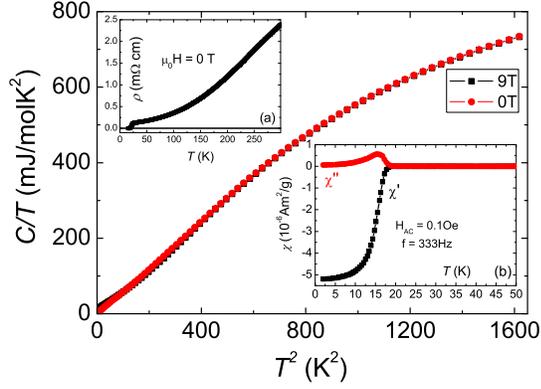}
\caption {(color online) Raw data of specific heat plotted as $C/T$
vs $T^2$ are shown in the main frame. The inset (a) and (b) show the
temperature dependence of the resistivity and the ac susceptibility
at zero dc magnetic field, respectively. } \label{fig1}
\end{figure}

The raw data of the specific heat in various magnetic fields at $T <
7.7$ K are shown in Fig. 2(a). One can see that the Schottky anomaly
is visible when $\mu_0 H \geq 2$ T, while it is negligible at lower
fields because the SH peak due to the Schottky anomaly still locates
in the lower temperature region (e.g. lower than 1.8 K) when the
field is low. So we just need to consider the correction from the
Schottky anomaly at high fields when fitting the data to separate
the electronic contribution from other contributions. As we all
know, the total specific heat of a superconductor can be expressed
as

\begin{equation}
C(T,H)=[\gamma_0+\gamma(H)]T+\beta T^3+C_{Sch}(T,H),\label{eq:1}
\end{equation}
where the four terms represent the residual electronic specific
heat, the electronic contribution, the phonon contribution and the
magnetic impurity contribution (the so-called Schottky anomaly),
respectively. The two-level Schottky anomaly is given by
$nx^2e^x/(1+e^x)^2$ $(x=g\mu_B H/k_B T)$ at nonzero fields, where
$\mu_B$ is the Bohr magneton and $n$ is the concentration of
paramagnetic centers. As revealed by the thick solid lines in Fig.
2(a), the data under the fields $\mu_0 H \geq 2$ T can be fitted
very well by equation (1). As for the data of lower fields, we just
simply extrapolated them down to 0 K linearly and consequently the
electronic contribution and the phonon contribution were separated.

\begin{figure}
\includegraphics[width=8cm]{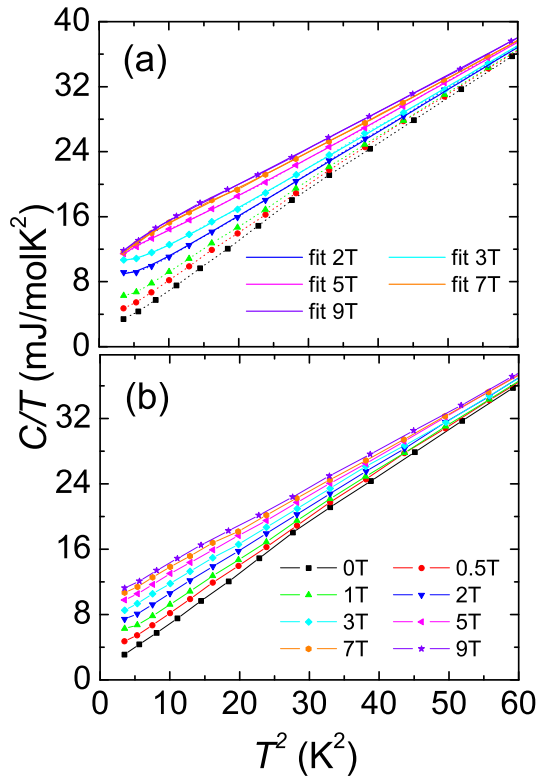}
\caption {(color online) Temperature and magnetic field dependence
of the specific heat in $C/T$ vs $T^2$ plot in the low temperature
region. (a) Raw data before removing the Schottky anomaly. We can
see that the Schottky anomaly becomes visible when the field is
higher than 2 T. The thick solid lines represent the theoretical fit
(see text) containing all terms in eq.(1). (b) Replot of the data
after the Schottky anomaly was subtracted. The dotted lines in (a)
and solid lines in (b) are guiding to the eyes.} \label{fig2}
\end{figure}

From the treatment of the data as described above, we obtained the
fitting parameters for different fields, as shown in Table-I. Here
we take the average value $\beta$ $\sim$ 0.49 mJ/(mol K$^4$). And
the residual term $\gamma_0$ was determined to be 0.69 mJ/(mol
K$^2$), which is an extremely small value compared with that in
other systems. The small contribution from Schottky anomaly and
residual term make our data analysis more easy and reliable. Using
the obtained value of $\beta$ and the relation $\Theta_D$ =
$(12\pi^4k_BN_AZ/5\beta)^{1/3}$, where $N_A$ = 6.02 $\times 10^{23}$
mol$^{-1}$ is the Avogadro constant, Z = 8 is the number of atoms in
one unit cell, we get the Debye temperature $\Theta_D $ = 315.7 K
for the present sample. Suppose it is in the weak electron-phonon
coupling regime, we then use the McMillan equation to evaluate the
electron-phonon coupling strength $\lambda_{e-ph}$
via\cite{McMillan}

\begin{equation}
T_c=\frac{\Theta_D}{1.45}\exp[-\frac{1.04(1+\lambda_{e-ph})}{\lambda_{e-ph}-\mu^*(1+0.62\lambda_{e-ph})}],
\end{equation}
where $\mu^*$ is the Coulomb pseudopotential taking about 0.10.
Using $\Theta_D$ = 315.7 K and $T_c$ = 26 K, we obtain a very large
value for the electron-phonon coupling: $\lambda_{e-ph}$ = 1.2. This
may suggest that a simple model based on electron-phonon coupling
and the McMillan equation cannot interpret the high transition
temperature in the present system.

\begin{table}
\caption{Fitting parameters of equation (1).}
\begin{tabular}{ccccc}
\hline \hline
$\mu_0H$(T) & $\gamma(H)$(mJ$/$mol$\,$K$^{2}$)   &  $\beta$(mJ$/$mol$\,$K$^{4}$)   & $n$(mJ$/$mol$\,$K)  & $g$ \\
\hline
0.0      & $0.000$       & $0.586$      & -        & -  \\
0.5      & $1.684$       & $0.584$      & -        & -  \\
1.0      & $2.914$       & $0.560$      & -        & -  \\
2.0      & $4.810$       & $0.522$      & $9.58$        & $2.01$  \\
3.0      & $6.026$       & $0.504$      & $9.70$        & $1.85$  \\
5.0      & $7.326$       & $0.486$      & $10.50$        & $2.18$  \\
7.0      & $8.310$       & $0.470$      & $11.68$        & $2.07$  \\
9.0      & $8.910$       & $0.466$      & $10.84$        & $1.79$  \\
 \hline \hline
\end{tabular}
\label{tab:table1}
\end{table}

In the following we investigate the magnetic field induced change of
the SH coefficient $\gamma(H)$. After removing the Schotkky anomaly
for each field, we are left with only the the first three terms in
eq.(1). The results are shown in Fig. 2(b). It is clear that the low
temperature part is quite straight for all fields, this allows to
determine the zero temperature value of $\gamma(H)$ at different
fields. As shown in Fig. 3, $\gamma(H)$ increases nonlinearly as the
magnetic field increases from 0 T to 9 T. In fact, the nonlinear
behavior can be roughly described by a simple equation $\gamma(H)$ =
A$\sqrt{H}$ as shown by the red solid line, which is actually the
theoretical prediction for superconductors with line nodes in the
gap function\cite{Volovik}. This suggests that in the 10\% F-doped
LaOFeAs sample the gap clearly has a nodal structure.

Although our data shows a relation being close to the d-wave
prediction $\gamma(H) \propto \sqrt{H}$, it is by no means to say
that the gap in the present sample is definitely of d-wave type,
since other type of pairing symmetry with nodes can also give rise
to a nonlinear $\gamma(H)$.  While we can nevertheless use the
d-wave model to derive some important parameters. For example, we
can obtain the maximum gap value by investigating the field-induced
term $\gamma(H)$ quantitatively, as we have done successfully in
LaSrCuO single crystals\cite{WenHH,YueW}. The term $\gamma(H)$,
which mainly arises from the Doppler shift of the quasi-particle
excitation spectrum near the nodes induced by the supercurrent
flowing around vortices, has a direct relation to the slope of the
gap at the node, $v_\Delta$=2$\Delta_0/\hbar k_F$ with $\Delta_0$
the d-wave maximum gap in the gap function $\Delta=\Delta_0$
cos$(2\phi)$, $k_F$ the Fermi vector( taking $k_F\approx\pi/a \sim$
0.78 ${\AA}^{-1}$, here $a=4.03\AA$ is the in-plane lattice
constant). We have known that the relation between $v_\Delta$ and
the prefactor $A$ is given by

\begin{equation}
A=\frac{4k_B^2}{3\hbar}\sqrt{\frac{\pi}{\Phi_0}}\frac{nV_{mol}}{d}\frac{a_L}{v_\Delta},\label{eq:3}
\end{equation}
where $\Phi_0$ is the flux quantum, $n$ is the number of conduction
planes per unit cell, $d$ is the c-axis lattice constant, $V_{mol}$
is the volume per mole, and $a_L=0.465$ for a triangular vortex
lattice\cite{Chiao,Vekhter}. The value of the prefactor $A$, which
is about 3.2 mJ/(mol K$^2$ T$^{0.5}$), has been obtained from
fitting the data in Fig. 3. So we can extract gap value $\Delta_0
\approx 3.4\pm0.5$ meV using the known parameters for our sample.
This is a reasonable value with the ratio
$2\Delta_0/k_BT_c\approx4.0$, if we take $T_c$ = 20 K. This ratio is
quite close to the prediction ($2\Delta_0/k_BT_c$ = 4.28) for the
weak coupling d-wave superconductors.

\begin{figure}
\includegraphics[width=8cm]{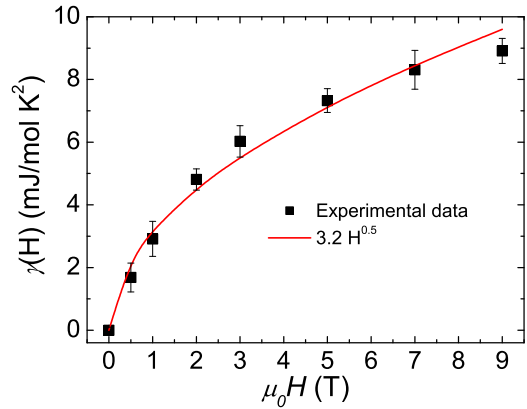}
\caption{(color online) Field dependence of the field-induced term
$\gamma(H)$ at $T = 0$ K (symbols). The solid red line is the fit to
$\gamma(H)$ = A$\sqrt{H}$. } \label{fig3}
\end{figure}

\begin{figure}
\includegraphics[width=8cm]{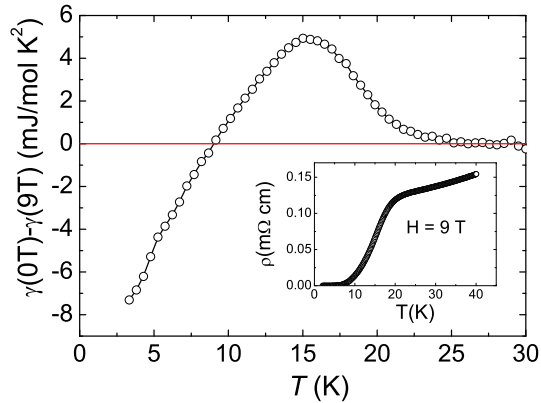}
\caption {(color online) Temperature dependence of
$\gamma(0T)-\gamma(9T)$ up to 30 K is shown in the main frame. One
can see a clear specific heat anomaly near $T_c$. The inset shows
the temperature dependence of resistivity at 9 T from 2 K to 40 K. }
\label{fig4}
\end{figure}

The nonlinear $\gamma(H)$ found here cannot be simply attributed to
the multigap effect as observed in $MgB_2$\cite{MgB2} and
$2H-NbSe_2$\cite{NbSe2} in which the $\gamma(H)$ is nonlinear. The
reason is that, even if there are multigaps in these two systems,
the zero field SH data $\gamma(T)$ shows a clear flattening in low
temperature region corresponding to the weak excitation of QPs for
an s-wave superconductor. This is completely absent in the present
system, as shown in Fig.1. One may further ascribe the nonlinearity
found for LaOFeAs to the granular feature of the sample: the grains
are randomly aligned within the bulk sample, when a field is
applied, the creating rates of DOS by a magnetic field are different
among the grains with different orientations. In this case, a
non-linear $\gamma(H)$ may be observed, especially in a system with
high anisotropy. This possibility cannot ruled out, but it is
difficult to understand why the relation $\gamma(H)\propto \sqrt{H}$
is roughly satisfied here.

In Fig.4 we present the temperature dependence of
$\gamma(0T)-\gamma(9T)$, one can see that a shallow SH anomaly
starts at about 25 K and ramps up slowly with decreasing
temperature, and shows a peak at about 15 K, which is just the
middle resistive transition point at 9T. This broadened anomaly may
be induced by both the broad transition at zero field (the
superconducting phase is still not perfectly uniform) and the very
low superfluid density leading to a strong phase fluctuation. Future
experiments on improved samples will fix these problems. Although a
magnetic field of 9 T is still difficult to suppress the
superconductivity completely, it is clear that the difference
between 9T and 0 T does not show a flattening down to about 2.8 K.
This is not expected by an s-wave BCS superconductor with $T_c$
beyond 20 K. This fact may also corroborate our conclusion derived
from the nonlinear behavior of $\gamma(H)$ at zero K that there are
nodes on the gap function.

In summary, the low temperature specific heat measurements reveal
that the new superconductor LaOFeAs has a rather low superfluid
density and condensation energy. The field induced extra DOS
$\gamma(H)$ follows a nonlinear behavior which is roughly
proportional to $\sqrt{H}$. This may suggest that LaOFeAs
superconductor has a nodal gap and is probably an unconventional
superconductor.

\begin{acknowledgments}
This work is supported by the National Science Foundation of China,
the Ministry of Science and Technology of China (973 project No:
2006CB601000, 2006CB921802), and Chinese Academy of Sciences
(Project ITSNEM).
\end{acknowledgments}


\begin{thebibliography}{00}

\bibitem{Tsuei1} C. C. Tsuei, and J. R. Kirtley, Rev. Mod. Phys. \textbf{72}, 969 (2000), and references therein.
\bibitem{Sigrist}M. Sigrist and K. Ueda, Rev. Mod. Phys. \textbf{63}, 239 (1991).
\bibitem{Kopnin1996} N. B. Kopnin and G. E. Volovik, JETP Lett. {\bf 64}, 690 (1996).
\bibitem{Volovik} G.E. Volovik, JETP Lett. {\bf 58}, 469 (1993); ibid 65, 491 (1997).
\bibitem{Moler} K. A. Moler, D. J. Baar, J. S. Urbach, Ruixing Liang, W. N. Hardy, and A. Kapitulnik, Phys. Rev. Lett.
\textbf{73}, 2744 (1994). K. A. Moler, John R. Kirtley, Ruixing
Liang, Douglas Bonn, and Walter N. Hardy, Phys. Rev. B \textbf{55},
12753 (1997).
\bibitem{Revaz} B. Revaz, J.-Y. Genoud, A. Junod, K. Neumaier, A. Erb, and E. Walker, Phys. Rev. Lett.
\textbf{80}, 3364 (1998).
\bibitem{Wright} D. A. Wright, J. P. Emerson, B. F. Woodfield, J. E. Gordon¡ì, R. A. Fisher, and N. E. Phillips, Phys. Rev. Lett.
\textbf{82}, 1550 (1999).
\bibitem{Phillips} N. E. Phillips, R. A. Fisher, A. Schilling, B. Buffeteau, T. E. Hargreaves, C. Marcenat, R. Calemczuk, A. S.
O'Connor, K. W. Dennis and R. W. McCallum, Physica B {\bf259-261},
546 (1999).
\bibitem{Nohara}M. Nohara, H. Suzuki, M. Isshiki, N. Mangkorntong, F. Sakai, and H. Takagi, J. Phys. Soc. Jpn. {\bf69}, 1602 (2000).
\bibitem{Chen}S. J. Chen, C. F. Chang, H. L. Tsay, H. D. Yang, J.-Y. Lin, Phys. Rev. B {\bf58}, R14753 (1998).
\bibitem{WenHH}H. H. Wen, Z. Y. Liu, F. Zhou, J. W. Xiong, W. X. Ti, T. Xiang, S. Komiya, X. F. Sun, and Y. Ando, Phys. Rev. B \textbf{70}, 214505 (2004).
H. H. Wen, L. Shan, X. G. Wen, Y. Wang, H. Gao, Z. Y. Liu, F. Zhou,
J. W. Xiong, and W. X. Ti, Phys. Rev. B \textbf{72}, 134507 (2005).
\bibitem{Hussey}For a review on the low energy quasiparticles, see N. E. Hussey, Advances in Physics 51, 1685 (2002).

\bibitem{LOMP1}P. Quebe, L. J. Terbuchte, and W. Jeitschko, J. Alloys Compd., {\bf302}, 72 (2000).
\bibitem{LOMP2}Y. Kamihara, H. Hiramatsu, M. Hirano, R. Kawamura, H. Yanagi, T. Kamiya, and H. Hosono, J. Am. Chem. Soc., {\bf128}, 10012 (2006).

\bibitem{LOMP3}T. Watanabe, H. Yanagi, T. Kamiya, Y. Kamihara, H. Hiramatsu, M. Hirano, and H. Hosono, Inorg. Chem. {\bf46}, 7719 (2007).
\bibitem{FirstJACS}Y. Kamihara, T. Watanabe, M. Hirano, and H. Hosono, J. Am. Chem. Soc. xxx, xxxx (2008).
\bibitem{McMillan}W. L. McMillan, Phys. Rev. \textbf{167}, 331 (1968).
\bibitem{YueW}Y. Wang, J. Yan, L. Shan, and H. H. Wen, Phys. Rev. B {\bf76}, 064512 (2007).
\bibitem{Chiao}M. Chiao, R. W. Hill, C. Lupien, L. Taillefer, P. Lambert, R. Gagnon, and P. Fournier, Phys. Rev. B {\bf62}, 3554 (2000).
\bibitem{Vekhter}I. Vekhter, P. J. Hirschfeld, and E. J. Nicol, Phys. Rev. B {\bf64}, 064513 (2001).
\bibitem{MgB2}F. Bouquet, et al., Phys. Rev. Lett. {\bf89}, 257001 (2002).
\bibitem{NbSe2}T. Hanagury et al., Physca B {\bf329}, 257001 (2002).





\end{thebibliography}
\end{document}